\documentclass[11pt,a4paper]{article}

 \newcommand{\be}{\begin{equation}}
 \newcommand{\ee}{\end{equation}}
 \newcommand{\bl}{\begin{equation}\begin{array}{ll}}
 \newcommand{\el}{\end{array}\end{equation}}
 \newcommand{\bll}{\begin{equation}\begin{array}{lll}}
 \newcommand{\bdm}{\begin{displaymath}}
 \newcommand{\edm}{\end{displaymath}}
 \def\bea{\begin{eqnarray}}
 \def\eea{\end{eqnarray}}
 \def\barr{\begin{array}}
 \def\earr{\end{array}}
 \newcommand{\bean}{\begin{eqnarray}}
 \newcommand{\eean}{\end{eqnarray}}
\def\p{\partial}
\def\d{\partial}

\def\f{\varphi}
\def\ve{\varepsilon}

\def\half{\frac{1}{2}}

\def\2third{\frac{2}{3}}
\def\4third{\frac{4}{3}}
\def\3quart{\frac{3}{4}}

\def\lim{\rightarrow}

\def\pr{\prime}

\def\bZ{\bar{Z}}

\def\bla{\bar{\lambda}}

\def\cL{{\cal L}}

\def\dop{\dot{p}}
\def\dq{\dot{q}}

\def\ddphi{\ddot{\f}}
\def\dchi{\dot{\chi}}

\def\dpsi{\dot{\psi}}
\def\dal{\dot{\alpha}}
\def\ddal{\ddot{\alpha}}
\def\dbe{\dot{\beta}}

\def\dga{\dot{\gamma}}

\def\deta{\dot{\eta}}

\def\dh{\dot{h}}

\def\Lag{{\cal L}}

\def\prbe{\beta^{\prime}}
\def\prga{\gamma^{\prime}}
\def\prpsi{\psi^{\prime}}
\def\prq{q^{\prime}}

\def\ttf{\tilde{\f}}

\begin{document}
\raggedbottom

\title{{\bf General properties and some solutions of generalized
Einstein - Eddington affine gravity I}}

\author{A.T.~Filippov \thanks{Alexandre.Filippov@jinr.ru}~ \\
{\small \it {$^+$ Joint Institute for Nuclear Research, Dubna, Moscow
Region RU-141980} }}

\maketitle

\begin{abstract}
 After a brief exposition of the simplest class
 of \emph{affine theories of gravity} in
 multidimensional space-times  with \emph{symmetric connections},
 we consider the spherical and cylindrical reductions of these
 theories to two-dimensional \emph{dilaton-vecton gravity} (DVG)
 field theories.
 The distinctive feature of these theories is the presence of a
 massive/tachyonic vector field (\emph{vecton}) with essentially
 nonlinear coupling to the dilaton gravity. In the massless limit, the
 classical DVG theory can be exactly solved for a rather general coupling
 depending only on the field tensor and the dilaton. We show that
 the vecton field can be consistently replaced
  by a new effectively massive scalar field
 (\emph{scalaron}) with an unusual coupling to dilaton gravity (DSG).

 Then we concentrate on considering the DVG models
 derived by reductions of $D=3$ and $D=4$ affine theories.
 In particular, we introduce the most general cylindrical reductions
 that are often ignored. The main subject of our study is the static
       solutions with horizons. We formulate the
 general conditions for the existence of the regular horizons and
 find the solutions of the static DVG/DSG near the horizons in the
 form of locally convergent power - series expansion. For an arbitrary
 regular horizon, we find a local generalization of the
 Szekeres - Kruskal coordinates.
 Finally, we consider one-dimensional integrable and nonintegrable DSG
 theories with one scalar. We analyze simplest models having three
 or two integrals of motion, respectively, and introduce the idea of
 a \emph{topological portrait} giving a unified qualitative description of
 static and cosmological solutions of some simple DSG models.

\end{abstract}
\section{Introduction}
    The present observational data strongly suggest that Einstein's
    gravity must be modified, one of the popular modifications being
    provided by superstring ideas. In view of the mathematical problems
    of the string theory, other, much simpler, modifications of gravity
    that affect only the gravitational sector (not touching other
    interactions) are also popular. One such modification was proposed and
    studied in \cite{ATF} - \cite{ATFg}.

   It is based on Einstein's idea (1923)\footnote{The references to his
   papers as well as to related work of Weyl and Eddington can be found
   in \cite{ATF} - \cite{ATFg}.} to formulate the gravity theory
   in a non - Riemannian space with a symmetric connection by use of
   a special variational principle that allows one to determine the connection
   from a `geometric' Lagrangian. This Lagrangian is  assumed to depend
   of the generalized Ricci curvature tensor and of other fundamental
   tensors and is varied in the connection coefficients.
   A new interpretation and a generalization
   of this approach was developed in \cite{ATFn} - \cite{ATFg} for an arbitrary
   space - time dimension $D$. In general, the resulting theory supplements
   the standard general relativity with dark energy (the
   cosmological constant,
   in the first approximation), neutral massive (or tachyonic) vector
   field (vecton) and (after a dimensional reduction to $D=4$) with $(D-4)$
    massive (or tachyonic) scalar fields.

   The further details of the theory depend on a concrete choice of
   the geometric Lagrangian, and then the corresponding physical theory can be
   described by the effective Lagrangian depending on the metric, vecton
   and scalar fields. In the geometric theory, there are no dimensional
   constants, while any fundamental tensor has dimension of some power
   of length (assuming $c=1$). We proposed a class of the geometric Lagrangian
   densities depending on the symmetric, $s_{ij}$,
   and antisymmetric, $a_{ij}$, parts of
   the generalized Ricci tensor as well as of a fundamental vector obtained
   by contracting the connection. Requiring the geometric `action'
   (i.e. the integral of the geometric Lagrangian density) to be dimensionless
   we can enumerate all possible actions. Thus, for $D=2n$ and $D=2n+1$
   we can construct $n+1$ independent scalar densities from the tensors
     $s_{ij}$ and $a_{ij}$
   having the weight two and dimension $L^{-D}$. Each Lagrangian is the square
   root of an arbitrary linear combination of these densities (of course,
   one can also take a linear combination of the square roots).

  The simplest useful density of this sort in any dimension is the square
 root of $\det(s_{ij} + \bla a_{ij})$, where $\bla$
 is a number.\footnote{Einstein used as the Lagrangian
 Eddington's scalar density
 $\sqrt{|\det r_{ij}|}\,$, where $r_{ij} \equiv s_{ij} + a_{ij}\,$.}.
 The effective physical Lagrangian is the sum of the standard Einstein
 term, the vecton mass  term,
 and the term proportional to $\det(g_{ij} + \lambda f_{ij})$
 to the power $\nu \equiv 1/(D-2)$, where  $g_{ij}$ and  $f_{ij}$
 are the metric
and the vecton field tensors (conjugate to $s_{ij}$ and $a_{ij}$),
 $\lambda$ is the number related to  $\bla$.
 The last term has the dimensional multiplier,
which in the limit of small field $f_{ij}$ produces the cosmological
constant. For $D=4$ we therefore have the term first introduced by
Einstein but now usually called the Born-Infeld or brane Lagrangian.
For $D=3$ we have the Einstein - Proca theory, which is very
interesting for studies of nontrivial space topologies. We will
 derive and study several analytical solutions of the
 one-dimensional reductions of this theory but their
 topological and physical meaning
 will be discussed elsewhere.

   A simplest dimensional reduction from $D>4$ to $D=4$ produces
   $(D-4)$ scalar fields which are also geometrically massive
   (or, tachyonic).
   The complete theory is very complex, even at the classical level.
   Its spherically symmetric sector is described by a much simpler
   1+1 dimensional dilaton gravity coupled to one massive vector
   and to several scalar fields. This dilaton gravity coupled
   to the vecton and massive scalars as well as its further
   reductions to 1+0/0+1 dimensional `cosmological' and `static'
   theories were first formulated in \cite{ATF} - \cite{ATFg},
   and here we begin systematic studies of their
   general properties and solutions. These studies are somewhat
   simplified by transforming the vecton field into a new massive scalar
   field, which is possible (on the mass shell in the 1+1
   dimensional reduction.\footnote{
   In addition to some formal simplifications, the transformation
   may shed a new light on some theoretical problems, especially, in
   cosmology. Indeed, new cosmological theories operate with
   several exotic scalar fields (inflatons, phantoms, tachyons), which
   are usually introduced \emph{ad hoc}. In the models considered
   here, such particles may appear in a consistent theoretical
   framework.}

  A class of exact solutions of the 1+1 dimensional dilaton-vecton gravity
  theory can be derived in the zero - mass case. We have shown that,
   if the mass of the vector field is zero and the scalar
 fields vanish, the dilaton-vecton gravity is explicitly integrable.
 This is true for any D, but the more or less explicit solutions
 are possible only for $D = 3,4,5$ (with arbitrary other parameters).
      One may hope that our solution can be used as a first approximation
 in a perturbative expansion and thus our approach opens a way to use
 a perturbation theory in the mass parameter. However, as can be
 seen from our treatment of the exact massive equations, such a
 perturbation must be rather nontrivial.
 In fact, as is demonstrated below, the 1+1 dimensional DVG, with a
 massive vector field, can be transformed into a dual DSG, with a
 scalar field replacing the vecton.\footnote{
 As distinct from the normal scalar fields, the scalaron has
 a different coupling to gravity and may have abnormal signs of
 the kinetic term (phantom) or mass term (tachyon).
 }
  This transformation allows us to
 use all general results obtained in recent years for
        the dilaton gravity coupled to the standard
 scalar fields. However, it shows that the perturbation theory may
 be `singular' because the equations of motion become degenerate
 in the limit of vanishing mass parameter (a well - known and more
 difficult problem of this sort is the so called boundary layer
 problem in hydrodynamics).

      After this brief overview of the results obtained in
 \cite{ATF} - \cite{ATFg} and some ideas of the present study,
 we give main equations which will be used below.
 Here we consider only the simplest {\bf geometric Lagrangian},
 \be
 \label{1.1}
 {\Lag}_\textrm{g} = \sqrt{-\det(s_{ij} + \bla a_{ij})}\,,
 \ee
 where the minus sign is taken because $\det(s_{ij})<0$
 (due to the local Lorentz invariance) and we
 naturally assume that the same is true for
 $\det(s_{ij} + \bla a_{ij})$
 (to reproduce Einstein's general relativity
       with the cosmological constant in the limit
 $\bla \rightarrow 0$).

 Following the steps of Ref.~\cite{ATFn}, we may write the
 corresponding {\bf physical Lagrangian}
  \be
 \label{1.2}
 \Lag_{\textrm{ph}} =  \sqrt{-g} \,
 \biggl[ -2 \Lambda \,[\det(\delta_i^j +
 \lambda f_i^j)]^{\nu} +  R(g) +
 m^2 \, g^{ij} a_i a_j \biggr] \,, \qquad
   \nu \equiv 1/(D-2) \,,
 \ee
  which should be varied with respect to the metric and the
  vector field; $m^2$ is a parameter depending on the
  chosen model for affine geometry and on $D$
  (see \cite{ATF} - \cite{ATFg}). This parameter can be
  positive or negative.
  In Einstein's first model it is negative.
     When the vecton field is zero,
  we have the standard Einstein gravity
  with the cosmological constant. Making the dimensional
  reduction from $D \geq 5$ to $D=4$,
  we obtain the Lagrangian describing
 the vecton $a_i$, $f_{ij} \sim \p_i a_j - \p_j a_i$
 and $(D-4)$ scalar fields $a_k, \, k = 4,..,D$.

 It is interesting to note that, for $D=3$, the Lagrangian
 (\ref{1.2}) is bilinear in the vecton field, and in the
 approximation $a_i = 0$ it gives the three-dimensional gravity
 with the cosmological constant. The three-dimensional gravity was
 studied by many authors\footnote{See, e.g., very interesting
 papers  of the last century, \cite{Star} - \cite{Welling}.
 Many ideas and results of these and other studies of the 3-gravity
 before 1997 were summarized in a beautiful review \cite{Carlip}
 (see also \cite{Banados}).  More recent studies of the three-dimensional
 gravity reveal new aspects of its relation to string theory and
 modern ideas on quantizing gravity, see, e.g. \cite{Das} - \cite{Witten1}.}
 and this may significantly simplify the study of solutions of the
 new theory. Unfortunately, we could not find works including into
 consideration both the cosmological constant and massive/tachyonic
 vectons and thus we begin our study by considering basic features of
 the simplest solutions of the theory not touching most interesting
 topological and quantum aspects.

 We first consider the simplest spherical dimensional reduction
 from the $D$-dimensional theory (\ref{1.2}) to the two-dimensional
 dilaton-vecton gravity (DVG) and then further
          reduce it to one-dimensional static
 or cosmological-type theories. The next step is considering
          cylindrical reductions of the $D$-dimensional theory
 to more complex
          two-dimensional dilaton gravity theories coupled to
          several scalar fields.
 For simplicity, we consider
 this dimensional reduction only in
          the dimensions four and three.
 As is well known, in the $D=4$ case, in addition to the dilaton,
 there appear `geometric' $\sigma$-model scalar fields and vector
 pure gauge fields that produce an effective potential
 introduced in \cite{ATFr}.
 These two-dimensional scalars look like scalar matter fields
 obtained by dimensional reductions from higher dimensional supergravity
 theory. Like those fields, they are classically massless.
 However, in general, the $\sigma$-model coupling of the scalar fields
 is supplemented by the mentioned effective potential depending on
 the scalar fields and on
 the `charges' of the `geometric' pure gauge fields. This theory is very
 complex and not integrable even after reduction to dimensions 1+0 or 0+1
 (see \cite{ATFr}). Of course, adding the vecton coupling makes the theory
 much more complex and in this paper we consider only the simplest
 variants.

 Much simpler, although quite nontrivial, is the three-dimensional
 theory (\ref{1.2}) (with $D=3$). We hope that its dimensional reductions
 to two-dimensional and one-dimensional DVG theories may give some
 insight into properties of the more realistic four-dimensional theory
 and thus study these models in more detail.
 For example, the cylindrical reductions are of great interest in
 connection with the theory of nonlinear gravitational
 waves\footnote{The first static cylindrical reductions were obtained in
 \cite{Tullio} - \cite{Weyl} and
 the first cylindrical nonlinear wave solutions were derived in
 \cite{ER}. Cylindrical solutions may be of interest for the theory
 of cosmic strings. They can also be related to static axially
 symmetric solutions like those derived in \cite{Ernst}.
 The cylindrical solutions in the presence of a massive vector field
 were not discussed in literature, to the best of our knowledge.}.

  In this paper we mainly discuss low-dimensional theories
 ($D=1,2,3,4$) that allow one to consistently treat some solutions
 of realistic higher dimensional theories.
   We mostly concentrate on mathematical problems and do not
 discuss in any detail
 dimensions and values of the physical parameters
 as well as physics meaning of the obtained solutions. Note also that
 in the context of modern ideas on inflation, multiverse, etc.,
 the main parameters of a fundamental theory of gravity cannot
 be theoretically determined. In particular, we do not know the sign
 and the magnitude of the cosmological constant (or any other parameter
 giving a fundamental scale of length). Even the dimension of the
 space - time (and, possibly, its signature) should be considered as
 a free parameter that can be estimated in the context of a concrete
 scenario of the multiverse evolution or by anthropic considerations.
 In practice, this means that we should regard theories with any
 parameters, in any space-time dimension as equally interesting,
 at least, theoretically.

 \section{Dimensional reductions of the generalized gravity theory}
 Let us outline the main reductions of the model (\ref{1.2}) in
 the dimensions $D=3, 4$.  Due to natural space and
 time restrictions we give only an overview of our \
 results\footnote{For a more rigorous discussion and
 applications of the reductions in $D=3,4$ see
 \cite{Chandra} - \cite{Grif}}.
 First, consider a rather general Lagrangian in the
 $D$-dimensional {\bf spherically symmetric} case:
 \be
 ds_D^2 = ds_2^2 + ds_{D-2}^2 =
 g_{ij}\, dx^i\, dx^j \,+ \,
 {\f}^{2\nu} \, d\Omega_{D-2}^2 (k) \, ,
  \label{2.1}
   \ee
  where $\nu \equiv (D-2)^{-1}$ and $k=0, \pm 1$.
  The standard spherical reduction
  of (\ref{1.2}) gives the effective Lagrangian
  \be
 \label{2.2}
 \Lag^{(2)}_D =  \sqrt{-g} \,  \biggl[\f R(g) +
 k_\nu \, \f^{1-2\nu} +
 {{1-\nu} \over {\f}} (\nabla \f)^2 +
  X(\f, \textbf{f}^{\,2}) - m^2 \f \, \textbf{a}^2
 \biggr] \,,
 \ee
 where $a_i(t,r)$ has only two nonvanishing components
 $a_0, a_1$, $f_{ij}$ has just one independent component
 $f_{01} = a_{0,1} - a_{1,0}$; the other notations are:
 $\textbf{a}^2 \equiv a_i a^i\,\equiv g^{ij} a_i a_j\,$,
 $\textbf{f}^{\,2} \equiv f_{ij} f^{ij}\,$,
 $k_\nu \equiv k(D-2)(D-3)$,  and, finally, \footnote{
       The expression for the determinant in Eq.(\ref{1.2})
       in terms of $f_{ij}$ for $D=4$ written in
       \cite{ATFn} contains also the fourth-order term
       the vecton field. It is not difficult to see that
       in both spherical and cylindrical reductions (see
       the end of this Section)
       this term vanishes and thus Eq.(\ref{2.3})
   is valid. In fact, this is also true in any dimension $D$.}
 \be
 \label{2.3}
 X(\f, \textbf{f}^{\,2}) \equiv
  -2 \Lambda \f \,\bigg[1 +
  \half \lambda^2 \textbf{f}^{\,2}\,\biggr]^{\,\nu} \,,
 \ee
    (here,
 the parameter $\lambda$ is related to $\lambda$ in
 (\ref{1.2}) but does not coincide with it in general).

 Sometimes, it is convenient to transform away the dilaton
 kinetic term by using the Weyl transformation, which in our
 case is the following:
 \be
 g_{ij} = \hat{g}_{ij} \, w^{-1}(\f) ,\,\,\,\,\,
 w(\f) = \f^{1-\nu}, \,\,\,\,\,\,
 \textbf{f}^{\,2} = w^2 \,\hat{\textbf{f}}^{\,2} \,,\,\,\,\,\,
 \textbf{a}^2 \,=\, w \, \hat{\textbf{a}\,}^2 \,.
 \label{2.4}
 \ee
 Applying this transformation to (\ref{2.2}) and omitting the
 hats we find the transformed Lagrangian
  \be
 \label{2.5}
 \Lag^{(2)}_{DW}  =  \sqrt{-g} \,  \biggl[\f R(g) +
 k_\nu \, \f^{-\nu} -
 2 \Lambda \f^{\nu} \,\bigg(1 +
  \half \lambda^2 \f^{2(1-\nu)}
  \textbf{f}^{\,2}\,\biggr)^{\,\nu}  -
  m^2 \f \, \textbf{a}^2 \biggr] \,.
 \ee
   When $D=3$ we have $\nu = 1$, $k_\nu =0$, Weyl's transformation
 is trivial and the Lagrangian is
  \be
 \Lag^{(2)}_{3}  =  \sqrt{-g} \,  \biggl[\f R(g) -
  2 \Lambda \f  -  \lambda^2 \Lambda \f \, \textbf{f}^{\,2} -
  m^2 \f \, \textbf{a}^2 \biggr] \,.
  \label{2.6}
 \ee
 These two-dimensional theories are essentially simpler than
 their parent higher dimensional theories.
 In particular, we show below
 that in dimension two the massive vector field theory
 can be transformed into a scalar dilaton gravity (DSG)
 model which is easier to analyze.
 Unfortunately, these DSG models and  their further
 reductions to dimension one (static and cosmological reductions)
 are also essentially non-integrable. It is well
 known that the massless case, being a pure dilaton gravity,
 is classically integrable even for an arbitrary coupling of
 the massless vecton to gravity (see, e.g., \cite{CAF} -
 \cite{Kummer} and reference therein, especially, in the review
 \cite{Kummer}). Having this in mind we try to find some
 additional integrals of motion.

 The next simplified theory is obtained in a
 {\bf cylindrically symmetric} case. We consider here only
 $D=3$ and $D=4$ cases. The general cylindrical reduction
 was discussed in detail in
 \cite{ATFr} and here we only summarize the main results.
 The most general cylindrical Lagrangian can be derived
 by applying the general Kaluza reduction to $D=4$. The
 corresponding metric may be written as
 \be
 ds_4^2 = (g_{ij} + \f \,\sigma_{mn} \,\f_i^m \f_j^n) \, dx^i dx^j +
  2 \f_{im} \, dx^i dy^m + \f \, \sigma_{mn} \, dy^m dy^n \, ,
 \label{2.7}
 \ee
 where $i,j = 0,1$, $m,n = 2,3$, all the
 metric coefficients depend only on the $x$-coordinates ($t,r$),
 and $y^m =(\phi, z)$ are coordinates on the two-dimensional
 cylinder (torus). Note that $\f$ plays the role of a dilaton and
 $\sigma_{mn}$ ($\det \sigma_{mn} = 1$) is the so - called
 $\sigma$-field.
 The reduction of the Einstein part of the four-dimensional
 Lagrangian, $\sqrt{-g_4} \, R_4 $, can be written as:
 \be
 \cL^{(2)}_{4 \textrm{c}}= \sqrt{-g} \, \biggl[ \f R(g) +
 {1 \over 2\f} (\nabla \f)^2  -
 {\f \over 4} {\rm tr} (\nabla \sigma \sigma^{-1}
 \nabla \sigma \sigma^{-1}) -
 {\f^2 \over 4} \sigma_{mn} \,\f^m_{ij} \,\f^{nij} \biggr] \, ,
  \label{2.8}
  \ee
 where $\f^m_{ij} \equiv \partial_i \f^m_j -
 \partial_j \f^n_i$.
 These Abelian gauge fields $\f_i^m$ are not propagating and their
 contribution is usually neglected. We proposed in \cite{ATFr}
 to take them into
 account by solving their equations of motion and writing the
 corresponding effective potential (similarly to what we are
 doing below in the spherically symmetric vecton gravity).
 Introducing a very convenient parametrization of the
 matrix $\sigma_{mn}$,
 \be
 \sigma_{22} = e^{\eta}\cosh\xi , \,\,\,\, \sigma_{33} =
 e^{-\eta} \cosh\xi, \,\,\,\, \sigma_{23} =
 \sigma_{32} = \sinh\xi \,,
  \label{2.9}
  \ee
 one can exclude the gauge fields $\f_i^m$ and derive the
 effective action
 \be
 {\cal L}_{\textrm{eff}}^{(2)} = \sqrt{-g} \, \biggl[ \f R(g) +
 {1 \over 2\f} (\nabla \f)^2  +
 V_{\rm eff}(\f, \xi , \eta) - {\f \over 2} [ (\nabla\xi)^2 +
 (\cosh \xi)^2  \,  (\nabla \eta)^2 ] \biggr] \, .
 \label{2.10}
 \ee
 where the effective geometric potential,
 \be
 V_{\rm eff} (\f, \xi , \eta) =
 -{\cosh \xi\over 2 \f^2} \biggl[Q_1^2 e^{-\eta}   -
 2 Q_1 Q_2 \tanh \xi + Q_2^2 e^{\eta}\biggr] \, ,
 \label{2.11}
 \ee
 depends on two arbitrary real  constants $Q_m$, which may be called
 `charges' of the Abelian geometric gauge fields $\f^m_{ij}\,$.

 This representation of the action is more convenient for writing
 the equations of motion, for further reductions to dimensions
 $(1+0)$, and $(0+1)$ as well as for analyzing special cases,
        such as $Q_1 Q_2 =0$, $\xi \eta \equiv 0$.\footnote{
        It is also closer to the
 original Einstein - Rosen equations for nonlinear
 gravitational waves \cite{ER}, which
 can be obtained by putting $Q_1 = Q_2 =0$ and $\xi \equiv                         0$.
      When $Q_1 Q_2 \neq 0$, $\xi$ and $\eta$ cannot be
      identically zero.}
      The  static solutions  of the theory (\ref{2.11})
      with $Q_1 Q_2 \neq 0$ have horizons while the exact solutions
      derived in  \cite{ATFr} for $Q_1 = Q_2 =0$ and
      nonvanishing $\sigma$-fields $\xi$, $\eta$
      have no horizons at all, in accordance with the general
       theorem of papers \cite{ATF1}.
      An interesting special case can be obtained if
      we choose $Q_1 = 0$,
      $Q_2 \neq 0$, $\xi \equiv 0$. In the Weyl frame
      the effective Lagrangian can be written in the form
    \be
 {\cal L}_W^{(2)} = \sqrt{-g} \, \biggl[ \f R(g)
  -{Q_1^2 \over 2\f^{5/2}} \, e^{-\eta}
  - {\f \over 2} \, (\nabla \eta)^2  \biggr] \, .
 \label{2.10a}
    \ee
   This is a standard dilaton gravity coupled to the
   scalar field $\eta$, with the potential depending
   both on the scalar field and dilaton. If $Q_1 \neq 0$,
   there exists a static solution with a horizon, which disappears
   in when $Q_1$ vanishes. Of course, the horizon exists also
   in pure dilaton gravity,  when $\eta \equiv 0$.
   In \cite{ATF1} we studied in some detail the models
   with the potentials independent of the scalar. Below we
   show that some results can be derived in some more
   general models with `separable' potentials
   $V(\phi, \psi) = v_1(\phi) v_2 (\eta)$.
   In particular, we show that one of the integrals of
   motion in the dilaton gravity coupled to massless scalars
   derived in \cite{ATF1} may exist also in some models with
   separable potentials that are of interest in the context
   of the present study.

  We see that the general cylindrical action is a very complex
  two-dimensional theory and even its one-dimensional reductions
  are rather complex and in general not integrable. Of course,
  adding the vecton sector does not make it simpler and more
  tractable. It deserves further studies mainly because it is
  much more realistic (from the physics point of view) than
  the spherically symmetric theory and still simpler than the
  the axially symmetric theory. Also, it is of physics interest
  because it can describe cosmic strings and in this direction one
  may hope to find some effects produced by the
  vecton coupling to gravity, i.e. traces of the affine
  geometry.

  Much more tractable is the \emph{three-dimensional} cylindrical
  space-time. The metric can be obtained by the obvious reduction
  of (\ref{2.7}),
  \be
 ds_3^2 = (g_{ij} +  \,\f_i\f_j) \, dx^i dx^j +
  2 \f_i \, dx^i dy + \f  \, dy^2 \, ,
 \label{2.12}
 \ee
 and the corresponding Einstein Lagrangian is simply
 \be
 \cL^{(2)}_{3\textrm{c}} = \sqrt{-g \,\,\f} \,
 \{ R(g)  - {\f \over 4} \,\,\f_{ij} \,\, \f^{ij} \} \,,
 \qquad  \f_{ij} \equiv \f_{i,j} - \f_{j.i} \,.
 \label{2.13}
 \ee
 Using the equation od motion for $\f_{ij}$ and introducing
 the corresponding effective potential (see \cite{ATFr}
 and more general derivation below) we derive the following
 two-dimensional dilaton gravity
 \be
 {\cal L}^{(2)}_{\textrm{eff}} = \sqrt{-g} \, \{ \phi R(g) -
 8 Q^2 \, \phi^{-3} \}\, , \qquad  \phi \equiv \sqrt{\f} \,.
 \label{2.14}
 \ee
 As distinct from the cylindrical reduction of the
  four-dimensional pure Einstein theory (corresponding
  to $Q=0$ in (\ref{2.14})), this theory has a horizon
  (see Section~4).

 It is not difficult to add the vecton part to the cylindrically
 symmetric Lagrangians. In fact the terms $-m^2 \f \, \textbf{a}^2 $,
 $X(\f, \textbf{f}^{\,2})$ (see (\ref{2.3})) are invariant and
 have the same form in any dimension. Therefore we can simply add
 the expressions (\ref{2.10}), (\ref{2.13}) to the gravitational
 Lagrangians also in the cylindrical case.
 However, there exist different cylindrical reductions of the vecton
 potential $a_i$. For example, unlike the spherical case,
 these fields may be nonzero for $i = 0,1,2$ and correspondingly
        $e_1 \equiv f_{01} \equiv \p_0 a_1 - \p_1 a_0 \neq 0$,
        $e_2 \equiv f_{02} \equiv \p_0 a_2 \neq 0$,
        $h_3 \equiv f_{12} \equiv \p_1 a_2 \neq 0$.\footnote{
        Here $e_i \equiv a_{0i}, \, h_i \equiv \ve_{ijk} a_{jk}$.
        In a diagonal metric $g_{ij} = g_i \delta_{ij}$ the
        fourth-order term in the determinant in Eq.(\ref{1.2})
        is proportional to
        $(e_i h_i)^2 \exp{\sum 2 g_k}$ and is seen to vanish.
        This argument obviously works in any dimension.
        }
              We see that the component $a_2 \equiv  a_{\f}$
       of the vector field $a_i$ behaves like an additional scalar
       field and, in addition to the two-dimensional vector field
       $(a_0, a_1)$ we have up to three scalar matter fields
       $a_2, \xi, \eta$ with a rather complex interaction to the
       dilaton gravity.
 This means that analyzing
  cylindrical solutions is more difficult than that of the
  spherical ones (see, e.g., \cite{Step} for
 different exact cylindrical solutions of Einstein - Maxwell
 theory). Our consideration of the cylindrical vecton solutions
 will be by necessity only fragmentary and superficial.
 In particular, in next Section we consider only the simple
 case of two potentials ($a_0, \,a_1$) and one field $f_{01}\,$.

\section{Nonlinear coupling of gauge fields and vector - scalar
duality}
    In the dimension $D=2$ all fields
 (vector, spinor, ...) are practically equivalent to scalar
 ones. Such equivalence is widely known for massless Abelian
 gauge fields (see, e.g., \cite{ATFHU}, \cite{VDATF1}
 and references therein).
 There exist also examples of such equivalence for
 some non - Abelian theories (see, e.g., \cite{Maison}). The aim of
 this Section is more modest -- to establish a standard map
 of (massive) Abelian vector fields to scalar fields.
 We do not attempt to find the most general result in this direction
 and restrict  our consideration to a rather general class of
 DVG coupled to some  scalar `matter' fields. This class includes
 all theories of the previous sections.

 Suppose that in place of the standard Abelian gauge field
 term, $X(\f,\psi) \textbf{f}^{\,2}$, the Lagrangian contains
 a more general coupling of
 the gauge field $f_{ij} = \p_i a_j -\p_j a_i$
 to dilaton and scalar fields, $X(\f, \psi; \textbf{f}^{\,2})$,
 where $\textbf{f}^{\,2} \equiv \,f_{ij} \,f^{ij}$.
 Using the Weyl transformation (if
 necessary) we may write a fairly general two-dimensional
 Lagrangian as
  \be
  {\cal L}^{(2)} = \sqrt{-g}\, \biggl[ \f R + V(\f, \psi) +
  X(\f, \psi; \textbf{f}^{\,2}) +  Z(\f)\, \textbf{a}^2 +
  \sum Z (\f ,\psi)(\nabla \psi)^2  \biggr] \,.
  \label{3.1}
  \ee
 We need not specify the number of the scalar matter fields and
 therefore
 omit the summation indices  for the matter fields $\psi$ and their
 $Z$-functions. For the fields, having positive kinetic
 energy, all the $Z$-functions
 in (\ref{3.1}) are negative and usually proportional to $\f$.
 We may call them `normal matter' fields or simply normal fields.
 This is not true for the
 dilaton $\f$ as seen in Eq.{\ref{2.2}; in particular the sign
 if the dilaton term can be changed and it can be even
 transformed to zero.

 In \cite{ATFHU} we considered the massless vector fields ($Z
 \equiv 0)$ and proposed to use instead of (\ref{3.1})
 the effective Lagrangian not containing the Abelian gauge fields
 (for a detailed proof see \cite{VDATF1}):
  \be
  {\cal L}_{\rm eff}^{(2)} = \sqrt{-g}\, \biggl[ \f R + V(\f, \psi) +
  X_{\rm eff}(\f, \psi; q) + \sum Z(\f ,\psi)(\nabla \psi)^2
  \biggr] .
  \label{3.2}
  \ee
 Here $q$ are integration constants (charges)
 defined by the solution of the equations for $f^{ij}$,
 \be
 2 \p_j \,(\sqrt{-g}\,X^{\prime} f^{ij}) =
 Z(\f) \sqrt{-g} \, a^{i} \,,
 \qquad X^{\prime} \equiv X^{\prime}(\f, \psi;
 \textbf{f}^{\,2}) \,
 \equiv {{\p X} \over  {\p \, \textbf{f}^{\,2}}} \,,
  \label{3.3}
 \ee
 which is useful to write in the LC coordinates $(u,v)$.
 Using the definitions and relations
 \be
 ds^2 = -4\,h(u,v)\, du\,dv \,, \quad  \sqrt{-g} = 2h \,,
 \quad f_{uv} \equiv a_{u,v} - a_{v,u} \,,
 \quad  -2\textbf{f}^{\,2} = (f_{uv} / h)^2 \,,
  \label{3.4}
 \ee
 we easily rewrite equations (\ref{3.3}) as
 \be
 \p_u (h^{-1} X^{\prime} f_{uv}) = - Z(\f) \, a_u \,, \qquad
 \p_v (h^{-1} X^{\prime} f_{uv}) =  Z(\f) \, a_v  \,.
 \label{3.5}
 \ee
 Defining  now the scalar fields $q(u,v)$ in the LC or in general
  coordinates by
 \be
  q(u,v) \equiv  h^{-1} X^{\prime} \, f_{uv} \,, \qquad
 2 \sqrt{-g} \, f^{ij} \, X^{\prime} \equiv
 \ve ^{ij} \, q \,, \quad i,j = 0,1 \,.
\label{3.6}
 \ee
 we see that they are constant in the massless case, when $Z = 0$,
 but in general  satisfy certain equations which we will derive
    in a moment. Then, Eqs.(\ref{3.5}) allow us to find
    the vector fields $\textbf{f}^{\,2}$ once we know
    the scalar fields $q$:
    \be
    a_u (u,v) = - Z^{-1} (\f) \, \p_u q (u,v) \,, \qquad
    a_v (u,v) = \, Z^{-1} (\f) \, \p_v q (u,v) \,.
   \label{3.5a}
    \ee

 Equations (\ref{3.6}) can be rewritten as equations for
 $\textbf{f}^{\,2}$ and can in principle be solved.
 Denoting the solution by
 $ \bar \textbf{f}^{\,2}$ (it depends on  $\f, \psi,  q$),
 we write them in the form
 \be
 2  \,\bar \textbf{f}^{\,2} \,=\, -(q / \bar X^{\prime})^2 \,, \qquad
 \bar X^{\prime} \,\equiv \,{\p \over \p \,{\bar \textbf{f}^{\,2}}}
 X(\f, \psi; \bar \textbf{f}^{\,2}) \,.
 \label{3.7}
 \ee
 Using this solution we can find different expressions for
 the {\bf effective action} that allows one to derive the
 explicit equations of motion for the fields $h, \f, \psi, q$.
 In \cite{VDATF1} we obtained two
 equivalent expressions for $X_{\rm eff}$\footnote{
 At this point, it is useful to recall that in the massless case
 $X = x(\f) \textbf{f}^{\,2}$ and thus
 $X_{\rm eff} = -x(\f)\bar \textbf{f}^{\,2} $, with
 $\bar \textbf{f}^{\,2} = q^2 /2 x^2(\f)$, where $q$ is a
 constant charge.
 }
 \be
 X_{\rm eff} (\f, \psi; q) \,=\,
  X(\f, \psi; \bar \textbf{f}^{\,2})
 -2  {\bar \textbf{f}^{\,2}} \bar X^{\prime}\, = \,
  X(\f, \psi; \bar \textbf{f}^{\,2}) +
  q^2 / \bar X^{\prime} \,.
  \label{3.8}
 \ee
  From (\ref{3.7}) and (\ref{3.8}) we also derived in \cite{VDATF1}
  the most compact and beautiful form for $X_{\rm eff}$:
  \be
  X_{\rm eff}(\f, \psi; q) =
  \biggl[ X(\f, \psi; \bar \textbf{f}^{\,2}) \,+\,
   q (u,v) \sqrt{-2 \,\bar \textbf{f}^{\,2}}\,\biggr]\,.
  \label{3.9}
 \ee
 It is a bit inconvenient because we have to correctly choose the
 signs of $q$ and $\sqrt{-2 \,\bar \textbf{f}^{\,2}}$ but in
 practice it is not difficult. The main difficulty in practical
 calculations is to explicitly derive $\bar \textbf{f}^{\,2}$
 as a function of $\f, \psi, q\,$ (though for $D=3,4$
 it is very easy).

 Now we can construct the effective Lagrangian giving all the
 equations of the new picture.
 Equations (\ref{3.3}) for $a_u \,, a_v$ obviously
 follow from (\ref{3.5}), (\ref{3.6}) that define the
 transformation. Calculating $f_{uv}$ by taking
 the partial derivatives of $a_u$ and $a_v$
 in Eq.(\ref{3.5a}) we find the equations of motion for
 the scalaron $q(u,v)$ in the standard form (see
 Eq.(\ref{4.5}) in next Section) but with somewhat unusual
 function $\bar{Z}(\f) = Z^{-1}(\f)$ as can be seen from
 (\ref{3.5a}).

      As in the case of a massless
     vector field we can simply replace $X$ by $X_{\rm eff}$
 For the proof of the on-mass-shell equivalence of the effective
 theory  (\ref{3.2}) supplemented by definitions and equations
 (\ref{3.3}) - (\ref{3.7}) and with $X_{\rm eff}$ given
 by (\ref{3.8}) or (\ref{3.9}) one can use the following easily
 checked identities:
  \be
 {d X_{\rm eff} \over dh} \,=\, 0 \,, \quad
 {d X_{\rm eff} \over d\f} \,=\,
 \p_{\f} \, X(\f, \psi; \bar \textbf{f}^{\,2}) \,, \quad
 {d X_{\rm eff} \over d\psi} \,=\,
 \p_{\psi} \, X(\f, \psi; \bar \textbf{f}^{\,2}) \,.
 \label{3.10}
 \ee
 Here we suppose that $q$ are independent variables
 and $\bar \textbf{f}^{\,2}$ are the functions of
 $\f, \psi, q$ satisfying equations (\ref{3.7}).
 To prove these relations we formally differentiate
 expression (\ref{3.9}) and use (\ref{3.7}). For example,
 \be
 {d X_{\rm eff} \over d\f} \,= \,\p_{\f} \, X \,+\,
  {d \bar \textbf{f}^{\,2} \over d\f}
 \, \biggl[\bar X^{\prime} -
 q / \sqrt{-2 \,\bar \textbf{f}^{\,2}}\,\biggr]
 = \,\p_{\f} \, X(\f, \psi; \bar \textbf{f}^{\,2}) \,,
 \label{3.11}
  \ee
 and the same vanishing  term in brackets
 emerges in
 the expression for the second and third terms of (\ref{3.10}).
 The first of identities in (\ref{3.10}) is a characteristic
 property of  $ X_{\rm eff}$ and thus can serve to define it
(see \cite{VDATF1}).

  Thus the solution of the DG coupled to scalar and
 Abelian gauge fields is reduced to solving DG coupled only
 to scalars $\psi$.  The special case, when
 $X(\f, \psi; \textbf{f}^{\,2}) =  X(\f) \textbf{f}^{\,2}$
 was known for long time and was used, for example,
 in finding charged spherical BH solution. The general theorem
 can be further generalized and applied to much more difficult
 problems. It was
   stated in \cite{VDATF1} that
  \emph {"It is not difficult to apply this construction to
 known Lagrangians of the Dirac - Born - Infeld type as well as to
 find new integrable models with nonlinear coupling of Abelian gauge
 fields to gravity."}. Now we realize this proposal by
 further generalizing this theorem that allows us
 to make a transformation of the neutral massive vector fields
 into neutral (and effectively massive) scalar fields with
 a somewhat unusual coupling to gravity.

 As a matter of fact, in the above construction we did not use the
 masslessness of $a$. And thus can try to define the
 effective potential $X_{\rm eff}$ for theory (\ref{3.1})
 by the same equations and definitions (\ref{3.3}) - (\ref{3.9}).
 Then, we consider equations (\ref{3.5}) as the expression
 of the vector field in terms of the new scalar field
 $q(u,v)\,$.\footnote{
 From now on we leave only one vector field and thus omit the
 subscript.} It is clear that the vecton mass term gives
 in the transformed Lagrangian
 the kinetic term of the scalaron,  while $X_{\rm eff}$
 defines the nonlinear coupling of the scalaron to DG.
 More generally, we expect that the two-dimensional
 DVG (\ref{3.3}) can be transformed into DSG
 \be
  {\cal L}_{\rm eff}^{(2)} = \sqrt{-g}\, \biggl[ \f R + V(\f, \psi) +
  X_{\rm eff}(\f; q(u,v)) + \bZ(\f)(\nabla q)^2 +
  \sum Z(\f ,\psi)(\nabla \psi)^2  \biggr] \,,
  \label{3.12}
  \ee
 which corresponds to a simplified version with one vector field
 generating one scalaron $q(u,v)$.
 For the spherically reduced vecton Lagrangian (\ref{2.5})
 with $D=4$
 we can easily derive,
 \be
   X_{\rm eff}(\f;\, q(u,v)) = -2\Lambda \sqrt{\f}
   \biggl[ 1 + q^2 / \lambda^2 \Lambda^2 \f^2  \biggr]^{\half} \,,
 \quad  V = 2k \f^{-\half} \,,    \quad \bZ = -1/m^2 \f \,,
   \label{3.13}
  \ee
 and there are no $\psi$-field terms.
 For the general cylindrical reduction we may  have two
 $\psi$-terms and $\psi$-dependent potential V (see (\ref{2.10})).
 For the $D=3$ case (\ref{2.6}) we have
 \be
   X_{\rm eff}(\f;\, q(u,v)) =
   -2\Lambda \f - q^2 / \lambda^2 \Lambda \f  \,,
 \quad  V = 0 \,,    \quad \bZ = -1/m^2 \f \,,
   \label{3.14}
  \ee

 Now, if we could solve the equations of motion for Lagrangian
 (\ref{3.12}) we would find the vecton field using
 the simple relation
 \be
 a_u =  \p_u \,q(u,v) / \,m^2 \f  \,, \qquad
 a_v = -\p_v \,q (u,v) / \,m^2 \f \,.
 \label{3.15}
 \ee
 We do not write here the general equations for DSG theory,
 the reader can find them in \cite{ATFr}, \cite{ATF1}.
 Most probably, the two-dimensional equations for $m^2 \neq 0$
 are not integrable. Even their one-dimensional reductions,
 described by ordinary differential equations
 seem to be not integrable. Below, we employ for their study
 some convergent or asymptotic expansions. To succeed with this
 one has to carefully study their general analytic properties
 and possible singularities.

 By the way, the `duality' between vecton and scalaron shows
 why a perturbation theory with the vecton mass as the parameter
 must be in some sense singular. In this paper we will not
 discuss it. The main subject of our study will be the
 two-dimensional DVG theory
 \be
  {\cal L}^{(2)} = \sqrt{-g}\, \biggl[ \f R + V(\f) +
  X(\f; \bar \textbf{f}^{\,2}) +
  Z_a(\phi) \, \textbf{a}^2 \biggr]\,,
  \label{3.16}
  \ee
 its dual DSG theory with $X_{\rm eff}(\f;\, q(u,v))$ given
 by (\ref{3.13}), (\ref{3.14}), and their reductions to dimension
 0+1 (static states). The most interesting solutions
 are those having local horizons which we roughly describe and
 classify in next Section. We first consider general properties
 of horizons and conditions for their existence for the general
 DSG theory given by (\ref{3.12}) with fairly arbitrary
 potentials. Then we discuss more concrete models and solutions
 with horizons. Our consideration could easily be adapted to
 cosmological models but we postpone this to near future.


\section{Horizons in DSG: a general theory}
\subsection{Equations}
 Consider a general two-dimensional DSG model that embraces all
 the above Lagrangians as well as some more general not yet
 discussed:
 \be
 {\cal L}_{\rm dsg} = \sqrt{-g}\, \biggl[ \f R + U(\f, \psi, q)
 + \bZ(\f)(\nabla q)^2 +  \sum Z(\f ,\psi)(\nabla \psi)^2  \biggr]
  \,.
  \label{4.1}
  \ee
 To simplify formulas, in the following we consider only one field
 $\psi$ and use the LC coordinates. It is not difficult to recover
 coordinate independent equations with any number of scalar fields
 (even with the `potentials' $Z(\f, \psi))$ depending on several
 fields $\psi$.
 Let us write the equations of motion in the LC coordinates
 (for more details see  \cite{ATF1},\cite{ATFr}). The energy and
 momentum constraints should be derived in general coordinates and
 then rewritten in the LC ones:
 \be
 h \,\p_i(\p_i \f / h) \,=\,
  \bZ (\f) \,(\p_i q)^2  +
  Z(\f ,\psi) \,(\p_i \psi)^2 , \qquad i=u,v.
 \label{4.2}
 \ee
 The equations of motion can be derived from the LC transformed
 Lagrangian,
  \be
  \half {\cal L}_{\rm dsg} = \f \, \p_u \p_v \ln|h| + h U(\f, \psi, q) -
  Z(\f, \psi) \, \p_u \psi \,\p_v \psi -
  \bZ (\f) \,\p_u q \,\p_v q \,,
  \label{4.3}
  \ee
 simply by variations in $\f$, $\psi$, $q$:
 \be
 \p_u \p_v \f + h \,U(\f, \psi, q) = 0 , \qquad
 \p_u (Z \,\p_v q) + \p_v (Z \, \p_u q) +
 h \,\p_q U(\f, \psi, q) = 0 ,
 \label{4.5}
 \ee
  \be
 \p_u (Z \,\p_v \psi) + \p_v (Z \,\p_u \psi) +
 h \,\p_{\psi} U(\f, \psi, q) =
 \p_{\psi} Z(\f, \psi) \,\p_u \psi \,\p_v \psi ,
 \label{4.6}
 \ee
 We omit the equation derived by variations in $\f$
 because it is satisfied as soon as equations (\ref{4.2}) and
 (\ref{4.5}) - (\ref{4.6}) are satisfied.\footnote{
 It simply gives the expression of the scalar two-dimensional
 curvature $R \equiv (\p_u \p_v \ln|h|)/h$ in terms of the
 $\f$ - derivative of the other terms in the Lagrangian. It
 may be useful in search for additional integrals,  e.g., \cite{ATF1}.}
 These equations can be
 reduced to dimensions 0+1 (static) or 1+0 (cosmological);
 the model can also describe some gravitational waves (see
 \cite{ATFr}, \cite{VDATF2}).

 In this paper we only consider
 static reductions, when all the fields are supposed to depend
 on one variable $u+v = \tau$.\footnote{We here use the notation
 $\tau$ instead  of $r$  because  $\tau$ does
 not coincide with the `radius' in  the
 static coordinates. We will see that a horizon can emerge when
 $\tau \rightarrow \infty$. In our picture the role of a `radius'
 plays the dilaton $\f$ which is finite at the horizon. The
 distinction between the space variable $r$ and time variable $t$
 can be established when we return to a higher dimensional origin
 of DSG. Note also that, due to the residual coordinate invariance,
 we can equivalently use transformed LC coordinates $a(u),\, b(v)$ in
 which $ds^2 = -4\,h \,a^\pr(u)\, b^\pr(v)$.}
 The most interesting solutions are those with horizons.
 We say that a solution has a horizon if the static metric
 $h(\tau)$ regarded as a function of $\f$ has a zero at a finite
 value of $\f$, i.e. $h \rightarrow 0$ for $\f \rightarrow \f_0$.
 Considering the Schwarzschild and Reissner - Nordstr{\o}m (S-RN)
 horizons\footnote{
 One can find a clear and concise LC treatment of black holes
 in \cite{Chandra}.}
 one can see that it is more convenient to replace the variable
 $\tau$ by $\f$ because then there will be no singularity at
 the horizon $\f = \f_{\,0}\,$. 
 For DSG theory (\ref{4.1}), it was shown
 in \cite{ATF1}, \cite{VDATF1}, \cite{FM} that we can rewrite the
  equations of motion for some new functions, which are finite
  and nonvanishing for $\f \rightarrow \f_{\,0}\,$. Moreover, these
  functions can be expanded in series of powers
  $\tilde{\f} \equiv \f -\f_0$ that converge in a neighbourhood
  of
   zero (provided that the potentials are analytic
  in this neighbourhood). This approach is applicable to a most
  general DG coupled to scalar fields but here we consider only
  the model (\ref{4.1}) with the scalaron and one extra scalar.

 Now, supposing that all the fields depend on $\tau$ and denoting
 the derivative in $\tau$ by the dot, we write the dynamical system
 corresponding to equations (\ref{4.2}), (\ref{4.5}) - (\ref{4.6}) in
 the form
  \be
 \dot{\varphi}\,\dot{h}/h + hU +\bZ \,{\dq}^2 + Z \,{\dpsi}^2 = 0\,;
 \qquad  \dchi +  hU  = 0\,,  \quad \dot{\varphi} = \chi \,,
 \label{4.7}
 \ee
 \be
  \bZ \dq = p \,, \quad
  2 \dop +  hU_q = 0  \,; \quad
  Z \dpsi = \eta \,, \quad
  2 \deta + hU_{\psi} = Z_{\psi} \,{\dpsi}^2\,,
  \label{4.8}
 \ee
 where $U_{\psi} \equiv \p_{\psi} U\,$, $Z_q \equiv \p_q Z\,$, etc.
 (in our model $\bZ_q = 0$ and in what follows we take $Z_{\psi} =0$).
 This first-order system is easy to rewrite in a Hamiltonian form,
 with the first equation in (\ref{4.7}) serving as the energy
 (Hamiltonian) constraint (see \cite{CAF}, \cite{ATF1}).
 However, representation (\ref{4.7}) - (\ref{4.8})
 (similar to what we used in \cite{ATF1}, \cite{VDATF1}, \cite{FM}
 but with the independent variable $\f$ instead of $\chi$)
 is more convenient for search and study of horizons.
 One of the reasons is the following. it is not difficult to show
 that $h \neq 0$ for finite $\tau$, and thus to find solutions near
 horizons we should work in asymptotic regions with, probably,
 asymptotic expansions. In addition, the distinction between
 regular and singular horizons discussed below
  is, at least, not evident.

 \subsection{Examples of horizons}
  Let us illustrate derivations of horizons by considering
  simplest  system (\ref{4.1}) with $U = U(\f)$,
  $q = q_0 = \textrm{const}$
  (massless vecton), and no extra scalars.
  Using (\ref{4.7}), with $q = q_0\,, \, \psi \equiv 0$, we find
 \bdm
 h  = C_0 \, \dot{\f} \,\,\,\mapsto  \, \ddphi +
 C_0 \,\dot{\f} \, U(\f) = 0 \,\,\,
 \mapsto \,\dot{\f} \,+\, C_0 \,N(\f) = C_1  \,,
 \edm
 where $U(\f) = U(\f, 0, q_0)\,$, $N(\f) \equiv \int U(\f) d \f $,
 $C_0$, $C_1 $ are constants.
 It follows
 \be
 h = C_0^2 \,[N_0 - N(\f)]\,, \qquad
 C_0 \tau = \int d\f \,[N_0 - N(\f)]^{-1}\,,
  \label{4.4}
  \ee
 where $N_0 = C_1 / C_0$ defines a finite position
   of the horizon in the interval
   $0 < \f_0 < \infty$ by the equation  $N(\f_0) = N_0\,$.
 As $N(\f)$ is continuous (and differentiable if $U(\f)$ is
 continuous), there exists at least one horizon
 for a generic potential. In the special cases of the
 Scwarzschild - Reissner-Nordstroem (S-RN)
  horizons Eq.(\ref{4.4}) gives the complete standard
 description of the black holes in any space-time
 dimension.\footnote{See, e.g., \cite{Chandra}, \cite{ATF1}
 and references therein. To get the standard formulas for the
 metric we must recall about the Weyl transformation
 of the metric (we use transformed Lagrangians (\ref{2.5}) instead
 of the original ones (\ref{2.2})), return to the space-time
 coordinates and take into account that $\f = r^{D-2}$).}
 The solution (\ref{4.4}) is valid for any potential $U(\f)$
 and describes a more general objects with any number of horizons
 that may be more complicated than than in the S-RN case.
  The S-RN black holes are the simplest examples of {\bf regular}
 and {\bf simple} (non - degenerate) horizons. In addition, the RN
 horizons may be double {\bf degenerate}, the corresponding black
 holes are then called extremal black holes.
 The potential describing these as well as somewhat more general
 black holes (with the  $\Lambda$ term)
 can be derived from the formulas of Section~3,
 \be
 U(\f) \,=\, k_{\nu} \,\f^{-\nu}  - 2\Lambda \,\f^{\nu} -
 q_0^2 \,\f^{\nu -2}\,, \qquad \nu \equiv 1/(D-2) \,.
  \label{4.4a}
  \ee
 It is not difficult to rewrite the S-RN solution (\ref{4.4})
 in the Schwarzschild, Eddington - Finkelstein or
 Szekeres - Kruskal (SK) coordinates. We will show how to
 construct a local analog of the solution (\ref{4.4}) for a
 general non-integrable theory (\ref{4.1}), demonstrate that
 all the mentioned types of horizons exist generally,
 and, by the way, introduce a local generalization of the
 SK coordinates.

 Before turning to this construction we discuss some general
 properties of the horizons described by Eq.(\ref{4.4}).
 If $N^\pr(\f_0) = 0$, the horizon is (double)
 {\bf degenerate}, i.e., it can be obtained by a fusion of
 two simple (nondegenerate) horizons.
  The additional condition for triple
 degeneracy of $\f_0$ is $U^\pr(\f_0) = 0$. This is possible
 if there are two relations between parameters determining the
 potential $U$, for (\ref{4.4a}) - between
 $\Lambda,\, q_0\,,\, k_{\nu}$. Therefore, in principle,
  there may exist in this case a triple degenerate horizon.
 If one of the parameters vanishes, only the double degeneracy is
 possible: 1.~$\Lambda\,= 0\,,\, k_{\nu} >0\,$;
 2.~$\Lambda\,<0\,,\, k_{\nu} =0\,$;
 3.~$\Lambda\, k_{\nu} >0\,$  for $q_{\,0} = 0\,$.
 It is not difficult to find that both conditions for
 the triple degenerate horizon can be
 satisfied if $\Lambda\,< 0\,,\, k_{\nu} >0\,$ and
 \bdm
  q_{\,0}^2 \,(2\Lambda)^{D-3} = (D-3)^{D - 3}\,(\nu \,k_{\nu})^{(D-2)}
  \equiv (D-3)^{2D - 5}\,,
 \edm
 where we used the definition of $k_{\nu}$.

 The  maximal number
 of possible horizons for an arbitrary potential $N(\f)$ can be
 derived by finding
 the  maximally degenerate horizon. The necessary conditions for
 this simple derivation are: the potential
 mast be sufficiently differentiable and continuously depending
 on the parameters, which can be deformed to make all the horizons
 to merge. Then the \textbf{number of possible horizons}
 coincides with the \textbf{maximal degeneracy} of them.
 A local generalization of this  `theorem' looks not so useful,
 being, at first sight, just a simple statement about splitting
 a degenerate horizon into simple ones.
 However, some global information could probably be extracted
 from more easily accessible local structure of horizons.
 We cannot say more at the moment but it would be useful
 to analyze this problem on different interesting examples.

 A very simple integrable model with a {\bf singular}
 horizon was given in \cite{FM}). This model  can be obtained
 from Eq.(\ref{4.7}) - (\ref{4.8}) by recalling that the first
 equation in (\ref{4.7}) is the energy condition stating
 that the Hamiltonian must vanish. Thus we can write the
 Hamiltonian
 \be
 {\cal H} = \dot{\varphi}\,\dot{h}/h + hU +\bZ \,{\dq}^2 +
 Z \,{\dpsi}^2 \,
 \label{4.4b}
 \ee
 introduce the canonical variables, and derive the canonical
 equations, which essentially coincide with (\ref{4.7})-(\ref{4.8})
 except for one extra equation giving  $(\ln h)\,\ddot{}\,$ in terms
 of the other variables (we omitted this equation above as it directly
 follows from the other ones). Now, suppose that $\bZ = -1$,
 $Z = 0$ and $U = g q^n$.\footnote{Hamiltonians of this sort
 can be met in simple approximations for branes (see, e.g.
 Cavaglia - Gregory).  As we work near the horizon $\f_0$
 we can approximate the dilaton field $\f$ by $\f_0$ in
 $\bZ$ and in $U$.} As ${\cal H}$ is independent of $\f$,
 we have $h = \exp(a + b \tau)$ and it is easy to find that our
 equations are integrable for $n = 1$ (trivial) and for
 $n = 2$ (not quite trivial but simple enough). In the $n = 1$
 case, the main result is
 \be
 q = q_{\,0} \,+\, h \,(g  /2 b^2)\,, \qquad \ttf =
 q_{\,0} h \,(g/b^2) -  h^2 (g^2 / 8 b^4) \,.
 \label{4.4c}
 \ee
 This equations define a regular simple horizon if $q_0 \neq 0$,
 but when $q_0 = 0$ we find $h \sim \pm \sqrt{|\ttf|}$.

 This example is important because it shows that singular horizons
 can be produced by completely regular potentials. Of course,
 if the potentials are not regular we expect that correspondingly
 there may exist singular horizons. To illustrate this by
 a simple and sufficiently realistic example we return for a moment
 to the pure DG theory the general solution of which is given by
 Eq.(\ref{4.4}). If $N(\f)$ is analytic for $\f > 0$, all
 horizons are regular. The points $\f_0$ where $N$ is nonanalytic
 may give a singular horizons. Consider the DG potential given
 by (\ref{3.14}), in which the $q$-field is a constant $q_0$ and
 take $\lambda^2 < 0\,$.\footnote{
 This is not unphysical assumptions for $\Lambda < 0$ because it
 corresponds to the massless vecton theory (\ref{2.2}) - (\ref{2.3})
 with the correct sign of the kinetic term
 $-\Lambda \lambda^2 \f \,\textbf{f}^{\,2}$ for small field values.}
 Neglecting the multipliers $\sim \f^{-\half}$
 that are analytic on $\f > 0$ we take the approximation
 preserving the singularity of the potential,
 \bdm
 U(\f) = -2\Lambda \sqrt{\f^2 - q_r^2} \,\equiv
 -2\Lambda \sqrt{\f_0^2 - q_r^2  + 2 \f_0 \,\ttf + \ttf^2}\,,
  \edm
 where $\f \equiv \f_0 + \ttf$ and $q_r$ is the renormalized
 $q_0$. Thus, for $\f_0^2 = q_r^{\,2}$ the potential is
 proportional to $\sqrt{\ttf}$ and $N(\f),\, h(\f)$ have
 the singularity  $\sim{\ttf}^{3/2}$.

 We believe that singular horizons must be regularly met
 in gravity but at the moment there are not many realistic
 examples.
 Probably, the reason is that
 all regular horizons (including the degenerate ones!) are
 similar in their construction but singular ones are
 singular each in their own way.

 \subsection{Solutions near horizons}
 Now we use as the independent variable $\f$ instead of $\tau$.
 Denoting the differentiation in $\f$ by the prime and defining
 $h/\chi \equiv H$, $p /\chi \equiv P$,   $\eta /\chi \equiv E$,
 we can transform the system (\ref{4.8}) into the following
 equivalent dynamical system ($Z_q = Z_{\psi} = 0\,$\footnote
 {This in not important for our argument but simplifies
 the expansions below. In a moment we further simplify the
 equations by neglecting the $\psi$-field which may emerge in
 cylindrical reductions.}):
 \bean
   q^\pr = P \bZ^{-1} \,, \quad
   \psi^\pr = E Z^{-1} \,,
 \quad (\chi P)^\pr = -\half H U_q \,,
 \quad (\chi E)^\pr = -\half H U_{\psi} \,,
 \nonumber\\
   \chi^\pr = -HU \,, \qquad \quad
  H^\pr = -H (P^2 \bZ^{-1} + E^2 Z^{-1}) \,. \qquad \qquad
 \label{4.9}
 \eean
 Suppose that $q(\f)$, $\psi(\f)$, $U(\f, q, \psi)$,
 $\bZ^{-1}(\f)$,
 and $Z^{-1}(\f)$ are finite when  $\ttf \rightarrow 0$
 and can be expanded in power series in $\tilde{\f}$, while
 $\, h /\tilde{\f} \, \rightarrow \, h_1 \, \neq \, 0\,, \infty\,$.
 It then follows that $H, E, P $ must be finite in this limit and
 can be expanded in power series. In particular, we find that
 $H \rightarrow H_0$ and from equations (\ref{4.9}) we have
 $\chi \rightarrow 0$,
 $P \rightarrow P_0\,$, $E \rightarrow E_0\,$, i.e. $p \rightarrow 0$,
 $\eta \rightarrow 0$. We thus can write
 \bean
 \quad
    h = \sum h_n\,\ttf^{\,n} \,, \qquad
 \chi = \sum \chi_n \,\ttf^{\,n} \,, \qquad
  q = \sum q_{\,n}\,\ttf^{\,n} \,, \qquad
  \psi = \sum \psi_n \,\ttf^{\,n} \,,
 \nonumber\\
 H = \sum H_n \,\ttf^{\,n} \,, \qquad
 P = \sum P_n \,\ttf^{\,n} \,, \qquad
 E = \sum E_n \,\ttf^{\,n} \,, \qquad \qquad
  \label{4.10}
 \eean
 where summation is over $0 \leq n < \infty$ and, of course,
 $h_0 = \chi_0 =0$.
 Using these expansions we can expand the potentials in powers
 of $\ttf$. To simplify notation we neglect the extra scalar
 and take $\psi = E = U_{\psi} = 0$. The expansion of $U,\, U_q\,,
 \,\bZ^{-1}$ can be explicitly derived by the expansions of
 $\f,\, q$ from (\ref{4.10}) (in our models
 (\ref{3.12}) - (\ref{3.14}) we have
 simply $\bZ^{-1} = -m^2 \f \equiv - m^2 (\f_0 + \ttf)$).
 To introduce notation we write a generic expansion of $X(\f, q)$
 and of $X(\f, q) Y(\f, q)$:
  \bean
 X(\f, q) = \sum  X^{(n)} \ttf^n \,,\qquad \quad
 X^{(0)} = X(\f_{\,0}, q_{\,0})\,, \quad
 X^{(1)} = X_{\f}^{(\,0)} + q_1 X_q^{(\,0)} \,,
 \nonumber\\
 X(\f, q)\, Y(\f, q) = \sum (X Y)^{(n)} \,\ttf^n \,,\qquad \quad
 (X Y)^{(n)} \equiv \sum_{m=0}^n (X)^{(n-m)} (Y)^{(m)} \,,
  \label{4.11}
 \eean
 where $X, Y$ may, in particular, coincide with one of the
 functions in (\ref{4.9}).

 With the assumptions formulated above, we can write the recurrence
 relations determining the coefficients of the power series
 (\ref{4.10}):
 \bean
 (n+1)\,\chi_{n+1} = - (UH)^{(n)} \,, \qquad \qquad
 2(n+1)\,(\chi P)^{(n+1)} = -(U_q H)^{(n)}\,, \quad
 \nonumber\\
 (n+1)\,q_{n+1} = (\bZ^{-1} P)^{(n)} \,\,,  \qquad \qquad
  (n+1)\,H_{n+1} = -(\bZ^{-1} P^2 H)^{(n)} \,.
 \quad
 \label{4.12}
 \eean
 The `initial' values  $ q_0, H_0$ and the position of the
 horizon, $\f_0$,  are arbitrary.
 Then we find
 \be
 \chi_1 = - U^{(0)} H_0 \,, \,\,\,
 \quad P_0 = U_q^{(0)}/2 U^{(0)}\,,
 \quad q_1 = \bZ_{(0)}^{-1} \,P_0 \,, \quad
 H_1 = - q_1 \,P_0^2 H_0 \,.
 \label{4.13}
 \ee
 Having this quadruple we can recursively derive any quadruple
 $\chi_{n+1}\,, P_n \,, q_{n+1} \,, H_{n+1}$ in terms of
 $\f_0 \,, q_0 \,, H_0$. It is not difficult to find several
 first quadruples although we do not know a general expression
 for any $n$.
  For further discussions
 we need $\chi_2 \,, h_2 \,$
 ( $h_1 = H_0 \,\chi_1$):
 \bean
 \chi_2 = - \half (U^{(0)} H_1 + U^{(1)} H_0) =
  - \half (U_{\f}^{(0)} + U^{(0)} \,P_0^2 \,\bZ_{(0)}^{-1})H_0\,,
  \nonumber\\
 h_2 \,=\, H_0 \,\chi_2 + H_1 \,\chi_1 =
  - \half (U_{\f}^{(0)} - U^{(0)} \,P_0^2 \,\bZ_{(0)}^{-1})H_0^2\,.
  \label{4.14}
 \eean

 Expansions  (\ref{4.10}) exist if the potentials do not vanish
 at $\f = \f_0$, i.e.,
 \be
 U^{(0)} \equiv U_{(0)} \equiv U(\f_0, q_0) \neq 0 \,, \qquad
 \bZ^{(0)} \equiv \bZ_{(0)} \equiv \bZ(\f_0, q_0) \neq 0 \,.
 \label{4.15}
 \ee
 For DSG theory (\ref{4.1}) (with $Z \equiv 0$), which is
 dual to  DVG theory (\ref{3.16}) and has  the effective
 potentials $U = X_{\rm eff} + V$ and $\bZ$ given by
  Eqs.(\ref{3.13}) - (\ref{3.14}), these
 conditions are satisfied if $\Lambda \neq 0$ and $0 < \f < \infty$.
 Then the power series solutions are well defined and converge
 near $\ttf = 0$ as argued in \cite{FM}.\footnote{
 If the first of conditions (\ref{4.15}) is not satisfied then
 to have a finite limit for $P_0$ the derivative  $U_q^{(0)}$
 must also vanish (see (\ref{4.13})). Then we see that
 $\chi_1 = h_1 = 0$ but $\chi_2$ and $h_2$ do not vanish. This is
 a characteristic property of (double) degenerate horizons.
 It is important that our expansion works also in
 this degenerate case. A general treatment of
 degenerate horizons will be published elsewhere.}.
 As we have seen above, there exist various
  obstructions to convergence: there can be singularities in the
  potentials similar to the discussed above;
  in addition, our nonlinear, non-integrable system can produce
  so called `moving singularities' depending on the initial values
  and parameters (like the critical dependence on
  $q_0$ in (\ref{4.4c})).\footnote{
  Here we should emphasize a peculiarity of our problem (\ref{4.9}).
  While we have four  unknown functions
  (with $\psi = E = U_{\psi} = 0$) and,
  at a first sight, a standard  Cauchy problem
  $\chi(0) = 0 ,\, H(0) = H_0,\, q(0) = q_0,\, P(0) = P_0\,$, in
  reality $P_0$  is expressed by (4.13) in terms of $q_0$ and
  $\f_0$ and thus cannot be arbitrary chosen (in our setting
  $\f_0$ is an arbitrary parameter of the system, not the initial
  condition).} A natural approach to singularities might be to
  find the leading singularity and try to look for corrections in
  the form of a convergent or asymptotic power series multiplier,
  like $\surd{\ttf} \sum a_n \ttf^{\,n}\,$. We will not pursue this
  idea further.

 \subsection{Generalization of Szekeres - Kruskal coordinates}
 In conclusion of this Section we introduce a generalization of the
 SK coordinates for any simple regular horizon for which the metric
 in (\ref{3.4}) is $\,h = h_1 \ttf + h_2 \ttf^2 + ...\,$, with $h_n$
 given by (\ref{4.12}) - (\ref{4.15}). As the metric changes sign
 with $\ttf\,$, from the physics point of view there is a `transition'
 of the static geometry to a time-dependent one. In the
 Schwarzschild coordinates, the metric is infinite at the horizon
 and also changes sign with $r-r_0\,$. This `singularity' disappears
 if one uses SK coordinates, the analog of which is easy to define
 by a simple  change of the chiral (LC) coordinates $u \mapsto a(u)$,
 $v \mapsto b(v)$:
 \be
 u = {\ln{a(u)} \over \chi_1}\,\,,\,\,\,
 v = {\ln{b(v)} \over \chi_1}\,\,; \qquad
 ab = \exp [\chi_1 (u+v)]\, \equiv \,\exp (\chi_1 \tau) \,=\,
 \exp \int d\f {\chi_1 \over \chi(\f)}\,.
 \label{4.16}
 \ee
 In view of the above expansions the last expression
 behaves for $\ttf \rightarrow 0$ like
 \be
 \exp \int d\f {\chi_1 \over \chi(\f)}\, = \,
 \ttf \exp \bigg[- {\chi_2 \over \chi_1} \,\ttf +
 \textrm{O}\,(\ttf^2)\biggr]\,,
 \label{4.17}
 \ee
 where $\ttf \equiv \f -\f_0$ and
 $\textrm{O}\,(\ttf^2)$ is a locally convergent power series.
 This means that the metric has no zero at  $\ttf \rightarrow 0$.
 Using (\ref{4.13}) - (\ref{4.17}) we rewrite the metric in
 $(a,b)$ coordinates:
 \be
 ds^2 = -4\,h(u+v)\, du\,dv
 \equiv -4\,h_{\textrm{sk}}(ab)\, da\,db =
  -4\,{h(\f) \over \chi_1^2 \,a b}\,\, da\,db
   \rightarrow  -4\,{da \,db \over U^{(0)}} \,\,.
 \label{4.18}
 \ee
 It is not difficult to derive first terms of the expansion of
 $h_{\textrm{sk}}(\ttf)$:
 \be
 h_{\textrm{sk}}(\ttf) = {h_1 \over \chi_1^2} \biggl[ 1 +
 \biggl({h_2 \over h_1} + {\chi_2 \over \chi_1}\biggr) \ttf
 + ...\biggr] =
 -{1 \over U^{(0)}}
 \biggl[ 1 + {U_{\f}^{(0)} \over U^{(0)}}\,\ttf + ...\biggr] \,.
 \label{4.19}
 \ee
 Interesting enough, these first terms depend only on $q_0$
 in $U^{(0)}$ and $U_{\f}^{(0)}$, not on $U_q^{(0)}$, which is
 proportional to $q_1\,$.\footnote{
 The first two terms in the expansion are insensitive to the
 parameters of the scalar fields and the expression for them
 coincide with the DG one. }
 Let us compare this result with the SK transformation for S-RN
 black holes. In this case we can write the closed expression for
 the transformed metric (see (\ref{4.4}) with $C_0 =1$):
 \be
 h_{\textrm{sk}}(\f) = {[N_0 - N(\f)] \over U_0^2}\,
 \exp\biggl(-U_0 \int d\f \,[N_0 - N(\f)]^{-1}\biggr)\,,
 \label{4.20}
 \ee
 where $U_{\,0} \equiv U(\f_{\,0})$, $N^\pr(\f) \equiv U(\f)$,
 etc. It is easy to find that the first approximation to this
 exact result coincides with Eq.(\ref{4.19}) (of course, the higher
 terms essentially depend on $q$).

 To compare our coordinates with SK ones, we briefly show
 how to rewrite our LC results in the standard Schwarzschild
 coordinates. First rewrite the metric in terms of coordinates
 $\tau = u+v \equiv r$ and $t = u-v$, so that
 $4 du dv = dr^2 -dt^2$ and thus
 \be
 ds^2 = -4h (\f)\,du \,dv  =
 -h(\f) \biggl[{d\f^2 \over \chi^2(\f)} - dt^2\biggr] \equiv
  -H(\f)\biggl[{d\f^2 \over \chi(\f)}  - \chi(\f)\, dt^2 \biggr] \,.
 \label{4.21}
 \ee
 To get the standard form of the static metric we must first to
 make the inverse Weyl transformation, i.e. to divide this
  metric by $w(\f) = \f^{1-\nu}$ (it is the same for
 the spherical and cylindrical reductions, see (\ref{2.4}))
 and to replace $\f = r^{D-2}$, $w = r^{D-3}$. Then we have the
 desired solution near a horizon in $(r,t)$ coordinates:
 \be
 ds^2 = -(D-2) \,H_s(r)\biggl[{dr^2 \over \chi_s(r)}  -
 \chi_s(r) \,dt^2 \biggr] \,; \qquad
 \chi_s(r) \equiv \nu \,\chi(r^{D-2})\,\, r^{3-D}\,,
   \label{4.22}
 \ee
 where  $H_s(r) \equiv H(r^{D-2})$. Reproducing the standard
 S-RN black hole solutions is now trivial: $H \equiv H_0 = C_0$
 and $\chi(\f)$ is given by Eq.(\ref{4.4}); deriving $N(\f)$
 for the potential (\ref{4.4a}) and adjusting notation we get
 the general spherical black hole metric in all the
 discussed coordinate: LC, Schwarzschild, and SK. Our most
 general solutions are only valid near horizons but, if we
 succeed in analytic continuation of them up to infinity
 $r^{D-2} = \f \rightarrow\infty$ or to singularity at
 the origin $\f=0 =r$,
 this solution may be regarded as global as S-RN ones.
 Hopefully, it may be possible for some solutions with
  maximally degenerate horizons, at least for special values
  of the parameters

 \section{Quasi summary and short remarks}
 The main results of this papers are the following.
 In Section~3 we proposed the transformation of the DVG into
 the equivalent DSG. This allows applying
 to the vecton models some methods
 developed in two-dimensional dilaton gravity models with
 scalar fields . In particular, the
 nonlinear kinetic terms of the vecton theory transform into
 completely standard potentials depending only on scalar fields
 (dilaton, scalaron, other scalars). The scalar form makes it
 easier to look for additional integrals of motion in the
 one-dimensional reductions of DSG
 (we may call them DSG1).\footnote{
 Pure dilaton gravity is a topological theory and thus  reduces
 to the one-dimensional integrable system. There exist simple enough
 examples of integrable systems involving one massless scalar field
 in addition to the dilaton (see \cite{ATF1}). More complex models
 with effectively massive scalar field may have one additional integral,
 at best, and thus remain nonintegrable. }
 In \cite{ATF1} (pp. 1698-1699) we derived two nontrivial DG models coupled
  to scalar fields and having two additional integrals. Using these integrals
  it was possible  to integrate them.
 We only mention here the first model of Ref.\cite{ATF1} that can be
 generalized to some effectively massive scalar scalar fields.
 It is not difficult to show that  for `multiplicative potentials
 $V = u(\f) v(\psi)$ there exists an additional integral of motion
 if the kinetic potential $Z_(\f)$ is related to $u(\f)$ as
 \be
  Z(\f) = (g_0 / u(\f)) \int u(\f) d\f \,,
   \ee
  while $v(\psi)$ is arbitrary function ($g_0$ is arbitrary constant).
  The integral is insensitive to $v(\psi)$ and is
  the same as given in \cite{ATF1} (for simplicity we choose
  there the Weyl frame in which $W=0$):
  \be
  Z h^{-1} \dh  -  g_1 \d{\f} = C_0 \,,
  \ee
  where $g_1$ is a constant depending on the integration constant in
  the definition of $Z$. This result may be of use in analyzing DSG
  introduced in Section~3.

  The second integrable model in \cite{ATF1} deserves special attention
  because it allows us to introduce a concept of \emph{the topological
  portrait} describing qualitative properties of static and
  cosmological solutions for different values of a parameter $\delta$
  characterizing the energy of the massless scalar field.
  Introducing the scales $w_0$ and $h_0$ for $w(\f)$ and $h(\f)$ we find
  the relation between normalized $w$ and $h$ that depends only on
  $\delta$ varying in the interval $[-\half , +\infty]$:
  \be
  w = {|h|^{\delta} \over |1 + \ve |h|^{1+2\delta}| } \,.
  \ee
 Here $\ve \equiv h/|h|$, $-1 < h < \infty \,, 0 < w < \infty$.
 It is not very difficult (but not quite easy) to draw the picture of
 the curves describing all possible solutions. In the domain  $h < 0$
 we have static  solutions, while for cosmological ones $h >0$.
 The picture looks like a phase portrait of a dynamical system in the $(h,w)$-plane,
   with  singular points: 0)~$(0,0)$,
 1)~$(0,1)$,  2)~$(-1, \infty)$, 3)~$(1, \half)$, 4)~$(0, \infty)$,
 5)~$(\infty, 0)$, 6)~$(-1,0)$ .
 These points are joined by the important separating curves.
 The most interesting points are: the node of the initial singularity, (p.0),
 the saddle point of the horizon, (p.1), and the cosmological point (p.3),
 at which all cosmologies tangentially coincide.

 While Section~4 presents a complete description of static solutions near
 horizons, the topological portrait, if available, will presumably allow
 us  to find the global picture and to make more clear the relation between
 static and cosmological solutions. It would be very helpful to find an
 approach to drawing 3D-portraits of integrable DSG systems.

 \section{Appendix}
 In this paper we use different coordinates in
 two-dimensional and one-dimensional dilaton theories.
 To make reading easier we give here short comments on the
 relation between the Lagrangians in the $(r,t)$ and $(u,v)$
 coordinates.  We first write metric (\ref{2.1}) in
 the $(r,t)$ coordinates,
 \be
 ds_{D}^2 = e^{2\alpha} dr^2 + e^{2\beta}
 d\Omega^2_{D-2} - e^{2\gamma} dt^2 + 2e^{2\delta} dr dt \, .
 \label{a.1}
 \ee
 Here the last term is needed to derive the momentum constraint,
 and while passing to the one-dimensional coordinates we
 omit it (formally, by $\delta \rightarrow -\infty$).
 Then action (\ref{2.2}) can be rewritten,
  \be
 \label{a.2}
 \Lag^{(2)}_D =
 e^{\alpha + (D-2) \beta + \gamma} \,
 \biggl[R^{(2)} + {1 - \nu \over \nu^2}
 [ e^{-2 \beta} + (\nabla \beta)^2]
 -2 \Lambda \,[1 +  \half \lambda^2 \textbf{f}^{\,2}\,]^{\,\nu}
   - m^2 \, \textbf{a}^2
 \biggr] \,,
 \ee
 where we omit the contribution of the non-diagonal term in
 metric (\ref{a.1}), because we use this expression only to
 connect one-dimensional reductions with DVG
 ($\nu \equiv (D-2)^{-1})$.
 The first two terms in (\ref{a.2}) belong to the standard
 Einstein gravity and the rest is the trace of the affine gravity.
 In the main text we also used the Lagrangians with
 additional scalar terms, like (\ref{3.1}) (one should keep
 in mind that its vecton part is the Weyl transformed version of
 Lagrangian (\ref{2.2})).

 Considering the expression for the curvature $R^{(2)}$
 in the diagonal  metric
  \be
  R^{(2)} = \,
  2[ e^{-2\gamma} (\ddal + \dal^2 - \dal \dga )
 - e^{-2\alpha} ({\gamma}^{\prime \prime} +
 {{\gamma}^{\prime}}^2 -  {\gamma}^{\prime} {\alpha}^{\prime})]\,,
 \label{a.3}
 \ee
 it is easy to find that the $R^{(2)}$-term in (\ref{a.2}) can be
 transformed into total derivatives and the
 terms containing only first-order derivatives.
 Neglecting the total derivatives we find that
 the remaining Einstein gravity
  terms  in (\ref{a.2}) have the form
    \be
 \label{a.4}
 \Lag^{(2)}_{DE} \equiv
  (D-2)(D-3)\, e^{\alpha + (D-2)\beta +  \gamma} \,
 \biggl[ e^{-2 \beta} \,+\,{\prbe}^2 -{\dbe}^2 +\,
 2 (D-3)^{-1} (\prbe \prga - \dbe \dga) \biggr] \,.
 \ee
 Using this expression it is easy to construct simple
 dimensional reductions to static or cosmological solutions
 (or employ more general methods of separation of variables,
 see \cite{ATFr}, \cite{VDATF2}). Of course we may apply
 the same approach to the Weyl transformed theory of
 DSG and rewrite Lagrangian (\ref{4.1}) in metric (\ref{a.1})
 as follows ($\f = \exp[(D-2)\beta]$):
 \be
 \label{a.5}
 {\cal L}_{\rm dsg} =
 \Lag^{(2)}_{DE} \,+\, e^{\alpha + \gamma}
 \biggl[ U(\f, \psi, q)
 + \bZ(\f)({\prq}^2 - {\dq}^2) \,+\
 \sum Z(\f ,\psi)({\prpsi}^2 - {\dpsi}^2) \biggr] \,.
 \ee
 This is a rather general formulation of dilaton gravity coupled
 to scalars, which is convenient to apply to many problems.
 A general approach to horizons was presented above.
  Many integrable cases were studied in literature,
  most relevant to the present paper are
  \cite{ATF1}, \cite{ATF2}, \cite{VDATF3}, \cite{VDATF2}.

 In Section~4 we derived one-dimensional equations
 directly from the two-dimensional LC equations
 (\ref{4.2}) - (\ref{4.6}). There, we could also dimensionally
 reduce Lagrangian (\ref{4.3}) and obtain the same
 one-dimensional equations.
 Alternatively, one may dimensionally reduce Lagrangian
 (\ref{a.5}) and obtain more general equations that allow
 gauge fixing. For example we can return to the LC gauge
 by choosing $\alpha = \gamma$ (this is possible also for
 the two-dimensional solutions). For static reductions
 we can choose the Schwarzschild gauge $\alpha(r) = -\gamma(r)$.
 In cosmological reductions a natural gauge is
 $\gamma(t) =0$. The LC equations are most suitable for studies of
 the states on both sides of horizons as well as for deriving
 solutions near horizons in general non - integrable theories.

 To avoid some small but annoying sign problems in relating
 the $(u,v)$ and $(r,t)$ pictures, let us explicitly write
 the conventions we are
 using in transitions between the pictures. We always use the
 following definition of the LC metric (often with $f(u,v)$
 instead of  $h(u,v)$
 \be
 ds_2^2 = e^{2\alpha} (dr^2 - dt^2) \equiv
 - 4 \,h(u,v) \, du \, dv \, .
 \label{a.6}
 \ee
 Then it is clear that the sign of $h$ must be positive for the
 space - like metric (this is true for the exterior space of
 the Schwarzschild black hole) and negative for the time - like
 one. For this reason, we choose the relation between the LC
 and space - time coordinates
 \be
 t = u + \ve v \,, \quad r = u - \ve v \,,
 \qquad \ve \equiv  |h|/h \, .
 \label{a.7}
 \ee
 With this definitions, $\p_t^2 - \p_r^2 = \ve \p_u \p_v$
 and the LC curvature is
 \be
 R^{(2)}_{LC} \,=\, 2 e^{-2\alpha}\,
 (\ddal - {\alpha}^{\prime \prime}) \,=\,
 h^{-1} \p_u \, \p_v \ln |h| \,, \quad
 \sqrt{-g} \, R^{(2)}_{LC} = 2 |h| R^{(2)}_{LC} \,=\,
 2\ve \p_u \, \p_v \ln |h| \,.
 \label{a.8}
 \ee
 Note that. if we take in all cases $\sqrt{-g} = 2h$
 instead of $2|h|$, the equations of motion  remain correct
 though the Lagrangians may change sign.
 Finally, in one-dimensional theories
 we usually take as the independent variable
 $\tau = u+v$, which is $r$ for static states and $t$
 for cosmologies.

 We also should mention simple definitions related to
 the vecton components:
 \be
 a_u = a_0 + a_1 \,, \quad  \ve a_v = a_0 - a_1 \,,
 \quad a_i a^i = - a_u a_v / h \,,
 \label{a.9}
 \ee
 \be
 f_{01} \equiv \,
 \p_0 \, a_1 - \p_1 a_0 = - \half \,\ve (a_{u,v} - a_{v,u})
 \equiv \half \,\ve f_{uv} \,.
 \label{a.10}
 \ee

 \bigskip
 {\bf Acknowledgment}

 This work was supported in part by the Russian Foundation for Basic
 Research:\\
 Grant No. 11-02-01335-a and Grant No.
 11-02-12232-ofi-M-2011.\\
 Useful remarks of E.A.~Davydov are kindly acknowledged.

 \bigskip
 
 \end{document}